\documentstyle[prl,aps,epsf]{revtex}
\newcommand{\be}{\begin{equation}}
\newcommand{\ee}{\end{equation}}

\begin{document}
\title{
Gravitating monopoles and black holes at intermediate Higgs masses}
\author{{\large Yves Brihaye} }
\address{\small 
Facult\'e des Sciences, Universit\'e de Mons-Hainaut,
B-7000 Mons, Belgium }
\author{and}
\address{ } 
\author{{\large Betti Hartmann, Jutta Kunz}}
\address{\small Fachbereich Physik, Universit\"at Oldenburg,
D-26111 Oldenburg, Germany }

\date{\today}

\maketitle
\begin{abstract}

Self-gravitating SU(2) Higgs magnetic monopoles
exist up to a critical value of
the ratio of the vector meson mass to the Planck mass,
which depends on the Higgs boson mass.
At the critical value a critical solution with a 
degenerate horizon is reached.
As pointed out by Lue and Weinberg,
there are two types of critical solutions,
with a transition at an intermediate Higgs boson mass.
Here we investigate this transition for black holes,
and reconsider it for the case of gravitating monopoles.

\end{abstract}

\section{Introduction}

In the SU(2)-Einstein-Yang-Mills-Higgs model,
gravitating magnetic monopoles and non-abelian black holes exist 
in a certain region of the parameter space \cite{lnw,bfm1,bfm2}.
For a fixed value of the Higgs boson mass,
gravitating monopoles 
exist up to a critical value $\alpha_{\rm cr}$ (of the parameter $\alpha$,
which is proportional to the ratio of the vector meson mass 
to the Planck mass), 
and non-abelian black holes
exist up to a critical value $\alpha_{\rm cr}(x_h)$ (for small enough
values of the horizon radius $x_h$ \cite{bfm2}).
At the critical value $\alpha_{\rm cr}$ a critical solution with a 
degenerate horizon is reached.
In particular, for small values of the Higgs boson mass,
the critical solution where a horizon first appears
corresponds to an extremal Reissner-Nordstr\o m (RN) solution
outside the horizon while it is non-singular inside.

Recently Lue and Weinberg \cite{lw} reconsidered
self-gravitating magnetic monopoles.
In particular, they observed, that 
for larger values of the Higgs boson mass,
the critical solution is an extremal black hole with non-abelian hair
and a mass less than the extremal RN value.
Exploring the transition between the two regimes,
occurring at some intermediate value of the Higgs boson mass,
Lue and Weinberg were left with a discrepancy between
their analytical and numerical results \cite{lw}.

Here we extend the investigation of Lue and Weinberg to 
non-abelian black holes,
showing the transition to persist in the presence of
fixed finite event horizon radii $x_h$.
Furthermore, we reconsider the case of gravitating monopoles
and argue, that the 
discrepancy between Lue and Weinberg's analytical and numerical
results \cite{lw}
can be traced back to their limited numerical analysis.

\section{Equations of Motion}

We consider the SU(2) Einstein-Yang-Mills-Higgs action,
with a Higgs triplet
\cite{lnw,bfm1,bfm2,lw,bhk,bhkt}
\begin{equation}
S= \int \frac{1}{16\pi G}R
\sqrt{-g} d^4x 
- \int 
 \left[ \frac{1}{4} F_{\mu\nu}^a F^{a\mu\nu} 
      + \frac{1}{2} D_\mu \phi^a D^\mu \phi^a
      + \frac{1}{4} \lambda (\phi^a \phi^a - v^2)^2 \right]
\sqrt{-g} d^4x
\ \label{action}   \end{equation}
where
\begin{equation}
F_{\mu\nu}^a = \partial_\mu A_\nu^a - \partial_\nu A_\mu^a
 + e \epsilon^{abc} A_\mu^b A_\nu^c
\ , \end{equation}
\begin{equation}
D_\mu \phi^a = \partial_\mu \phi^a + e \epsilon^{abc} A_\mu^b \phi^c
\ , \end{equation}
$e$ is the gauge coupling constant, $\lambda$ is the Higgs coupling
constant and $v$ is the Higgs field vacuum expectation value.
Variation of the action eq.~(\ref{action}) with respect to the metric
$g_{\mu\nu}$, the gauge field $A_\mu^a$ and the Higgs field $\phi^a$
leads to the Einstein equations and the matter field equations.
 
To construct static spherically symmetric globally regular
and black hole solutions
we employ Schwarz\-schild-like coordinates and adopt
the spherically symmetric metric 
\cite{bhk,bhkt}
\begin{equation}
ds^2=g_{\mu\nu}dx^\mu dx^\nu=
  -A^2N dt^2 + N^{-1} dr^2 + r^2 (d\theta^2 + \sin^2\theta d\phi^2)
\ , \end{equation}
with the metric functions $A(r)$ and $N(r)$,
\begin{equation}
N(r)=1-\frac{2m(r)}{r}
\ , \end{equation}
where $m(r)$ ist the mass function.

For the gauge and Higgs fields we employ the
standard spherically symmetric ansatz 
with vanishing time component of the gauge field
\cite{lnw,bfm1,bfm2,lw} 
\begin{equation}
\vec A_t = 0 \ , \ \ \
\vec A_r=0 \ , \ \ \
\vec A_\theta =  -\vec e_\phi \frac{1- K(r)}{e} \ , \ \ \
\vec A_\phi =   \vec e_\theta \frac{1- K(r)}{e} \sin \theta
\ , \end{equation}
and 
\begin{equation}
\vec \phi = \vec e_r H(r) v
\ , \end{equation}
with unit vectors $\vec e_r$, $\vec e_\theta$ and $\vec e_\phi$.

We now introduce the dimensionless coordinate $x$
and the dimensionless mass function $\mu$,
\begin{equation}
x = e v r \ , \ \ \ \mu=e v m
\ , \label{xm} \end{equation}
as well as the coupling constants $\alpha$ and $\beta$,
\begin{equation}
\alpha^2 = 4 \pi G v^2 \ , \ \ \ \beta^2 = \frac{\lambda}{g^2}
\ . \label{albe} \end{equation}

The $tt$ and $rr$ components of the Einstein equations then yield
the equations for the metric functions,
\begin{equation}
\mu'=\alpha^2 \Biggl( 
     N K'^2 + \frac{1}{2} N x^2 H'^2
   + \frac{(K^2-1)^2}{2 x^2} + H^2 K^2
   + \frac{\beta^2}{4} x^2 (H^2-1)^2 \Biggr)
\ , \label{eqmu} \end{equation}
and
\begin{eqnarray}
 A'&=&\alpha^2 x \Biggl(
    \frac{2 K'^2}{x^2} + H'^2 \Biggr) A
\ , \label{eqa} \end{eqnarray}
where the prime indicates the derivative with respect to $x$.
For the matter functions we obtain the equations
\begin{eqnarray}
(A N K')' = A K \left( \frac{K^2-1}{x^2} + H^2
 \right)
\ , \end{eqnarray}
and
\begin{eqnarray}
( x^2 A N H')' = A H \left( 2 K^2 + \beta^2 x^2 (H^2-1) \right)
\ . \label{eqH} \end{eqnarray}

The equations of motion depend only on the two physical parameters
$\alpha$ and $\beta$, eq.~(\ref{albe}).
They are related to the parameters $a$ and $b$
employed by Lue and Weinberg \cite{lw} via
\be
     a = 2 \alpha^2 = 8 \pi G v^2 \ \ \ , \ \ \ 
     b = \frac{1}{2} \beta^2 = \left(\frac{m_H}{2 m_W}\right)^2 \ .
\ee

We note that the embedded RN solution
with mass $\mu_\infty$ and
unit magnetic charge 
\begin{equation}
\mu(x) = \mu_\infty - \frac{\alpha^2 }{2x} \ , \ \ \ A(x)=1 \ , \ \ \
K(x)=0 \ , \ \ \
H(x)=1 
\ . \label{RN1} \end{equation}
is a special solution of these equations.
The extremal RN solution, in particular, has horizon $x_h$,
\begin{equation}
x_h = x_0 = \mu_\infty  = \alpha 
\ . \label{RN2} \end{equation}

\section{Gravitating Monopoles}

Let us first consider the globally regular particle-like solutions
of the SU(2) EYMH system, the gravitating monopoles.
Requiring asymptotically flat solutions implies
that the metric functions $A$ and $\mu$ both
approach a constant at infinity.
We here adopt
\begin{equation}
A(\infty)=1
\ , \label{bc1} \end{equation}
and $\mu(\infty)=\mu_\infty$ represents the dimensionless mass
of the solutions.
The matter functions also approach constants asymptotically,
\begin{equation}
K(\infty)=0 \ , \ \ \ H(\infty) = 1
\ . \label{bc2} \end{equation}
Regularity of the solutions at the origin requires 
\begin{equation}
\mu(0)=0 \ , \ \ \ K(0)=1 \ , \ \ \ H(0)=0
\ . \label{bc0} \end{equation}

Let us now recall how the gravitating magnetic monopole solutions
approach critical solutions,
when the vector boson mass, i.e.~$a$, is varied,
while the ratio of the Higgs boson mass
to the vector boson mass, i.e.~$b$, is kept fixed.
We focus on intermediate values of $b$.

Lue and Weinberg \cite{lw} realized that there are two
regimes of $b$, each with its own type of critical solution.
In the first regime $b$ is small. 
Here the metric function $N(x)$ 
of the monopole solutions always possesses a single minimum.
With increasing $a$
the minimum of the function $N(x)$ decreases.
For $a \rightarrow a_{\rm cr}$,
the critical solution is approached,
where the minimum of the function $N(x)$
reaches zero at $x = x_0$, eq.~(\ref{RN2}).
For $x \ge x_0$
the critical solution
corresponds to an extremal RN black hole solution with horizon radius
$x_h = x_0$ and unit magnetic charge, eq.~(\ref{RN1}).
Consequently, the mass of the limiting solution
coincides with the mass of this extremal RN black hole, eq.~(\ref{RN2}).
However,
in the interior region, $x < x_0$,
the critical solution is not singular,
in particular $N(0)=1$.
We refer to this approach to the critical solution as RN-type behaviour.
RN-type behaviour was seen for gravitating monopoles
and black holes \cite{lnw,bfm1,bfm2} 
as well as for gravitating dyons and dyonic black holes
\cite{bhk,bhkt}.

In the newly found \cite{lw} second regime $b$ is large,
and the metric function $N(x)$ 
of the monopole solutions develops a second minimum
as the critical solution is approached.
This second minimum arises interior to the location of the first minimum,
the solution therefore now possesses an inner and an outer minimum.
Once present, the inner minimum decreases faster 
with increasing $a$ than the outer minimum.
Consequently,
the inner minimum reaches zero, when the outer minimum 
still has a finite value, which is indeed not too different from the value
it had, when the inner minimum first appeared.
Thus for $a \rightarrow a_{\rm cr}$,
the inner minimum of $N(x)$ reaches zero at $x=x_* < x_0$.
This second approach to the critical solution
is demonstrated in Fig.~1,
where we show the function $N(x)$ 
for $b=36.125$ and several values of $a$ close to $a_{\rm cr}$.
Here the critical solution possesses an extremal horizon
at $x_* < x_0$,
corresponding to an extremal black hole with non-abelian hair
and a mass less than the extremal RN value.
Consequently, we refer to
this approach to the critical solution as 
NA-type (non-abelian-type) behaviour.

In their paper \cite{lw} 
Lue and Weinberg investigated analytically
the transition from RN-type behaviour to NA-type behaviour.
One of their central analytical results is, that RN-type behaviour
is possible only for $a > 1.5$,
making $a_{\rm tr}=1.5$ the critical value, where the transition
from RN-type to NA-type behaviour should occur.
This lower bound is found by an accurate analytical expansion
of the solution about the point $x = x_0$,
where $N(x)$ has a double zero.
The hypotheses used are mild and, very likely, 
fulfilled by the solutions.

When investigating the transition
from RN-type to NA-type behaviour numerically,
Lue and Weinberg \cite{lw} 
observe, that
with increasing $b$ the critical value $a_{\rm cr}$ decreases.
For small $b$ the critical value $a_{\rm cr}$ is larger than
the analytical transition value $a_{\rm tr}$,
and the behaviour is of RN-type as expected.
As $b$ is increased, and $a_{\rm cr}$ decreases towards
$a_{\rm tr}=1.5$ the behaviour remains of RN-type.
($a_{\rm tr}=1.5$ would correspond to $b \approx 25.6$ \cite{lw}.)
But surprisingly, as $a_{\rm tr}=1.5$ is passed,
Lue and Weinberg do not immediately observe the transition to 
NA-type behaviour.
Instead they continue to see numerically RN-type behaviour
until a value $a_{\rm cr}=1.42$,
corresponding to $b \approx 40$,
which is clearly below their analytical prediction for the transition.

Convinced by the analytical argument given by Lue and Weinberg
\cite{lw} for the transition value,
we argue in the following that the numerically observed discrepancy
arises only because of their interpretation
of their in accuracy limited numerical analysis.

Performing an independent numerical analysis (see Appendix for details),
we observe that NA-type behaviour occurs indeed
for lower values of $b$ than $b=40$ and thus
larger values of $a$ than $a=1.42$.
Numerically we find
NA-type behaviour for $b > 30 $.
This is illustrated in Figs.~2 and 3,
where we present 
the critical value $a_{\rm cr}-1$ 
and the ratio $x_*/x_0$ as functions of $b$,
respectively,
in the NA-type regime.
These figures should be compared to Figs.~10 and 8 of \cite{lw},
respectively.

A crucial point of the numerical analysis is illustrated in Fig.~4.
When for a fixed value of $b$ in the NA-type regime
the critical solution is approached,
the inner minimum appears at a certain minimal value of $a$.
In Fig.~4 we present
the value of the function $N(x)$ at the outer minimum 
at this certain minimal value of $a$
(where the inner minimum appears)
as a function of $b$.
Clearly, for $40 > b > 30$,
the inner minimum of the metric function $N(x)$ appears only,
when the value of the function $N(x)$ at the outer minimum 
is already very small.
Indeed, with decreasing $b$ it is decreasing 
from a value on the order of $10^{-6}$ to 
a value on the order of $10^{-9}$.
Obviously, with decreasing $b$ it becomes 
numerically increasingly difficult,
to observe NA-type behaviour at all.
As illustrated in Fig.~5 for $b=36.125$,
once the inner minimum is present,
the value of the inner minimum of $N(x)$
decreases rapidly with increasing $a$,
whereas the value of the outer minimum remains practically unchanged.

Even though our numerical analysis is not accurate
enough to find NA-type behaviour for $b<30$,
we expect it to be present until $a_{\rm tr}=1.5$ is reached.
Linear extrapolation of the curve in Fig.~2 leads to
$b_{\rm tr} = 25.58$,
while (a less accurate) extrapolation of the curve in Fig.~4 leads to
$b_{\rm tr} = 26.7$ \cite{foot}.
Thus our numerical analysis strongly supports the
analytical result of Lue and Weinberg \cite{lw}.

What concerns their numerical results,
we fully agree with their explicit calculations
as far as presented in \cite{lw}.
However, Lue and Weinberg do not consider numerical solutions,
where the value of the outer minimum is smaller than
$10^{-6}$ \cite{lw,lwp}.
Thus they miss
NA-type solutions for values of $b$ smaller than about 40,
as seen from Fig.~4. 
Indeed, if we extract from our numerical data a 
modified ``critical'' value of $a$
by requiring that the value of the minimum of $N(x)$
should be $10^{-6}$ \cite{lwp}
instead of requiring that it should be zero,
we obtain full agreement with Fig.~10 of Lue and Weinberg
\cite{lw}.

\section{Black holes}

We now turn to the black hole solutions of the
SU(2) EYMH system.
Imposing again the condition of asymptotic flatness,
the black hole solutions satisfy the same
boundary conditions at infinity 
as the regular solutions.
The existence of a regular event horizon at $x_h$
requires
\begin{equation}
\mu(x_h)= \frac{x_h}{2}
\ , \label{bc4} \end{equation}
and $A(x_h) < \infty $,
and the matter functions must satisfy
\begin{eqnarray}
 {N' K' }|_{x_h} =  \left. K \left( \frac{K^2-1}{x^2} + H^2 
 \right) \right|_{x_h}
\ , \label{bc5} \end{eqnarray}
and
\begin{eqnarray}
 {x^2  N' H' }|_{x_h} =  \left. 
   H \left( 2 K^2 + \beta^2 x^2 (H^2-1) \right)
  \right|_{x_h}
\ . \label{bc6} \end{eqnarray}

Magnetically charged non-abelian black hole solutions
exist only in a limited region of the $(a,x_h)$ plane \cite{lnw,bfm1,bfm2},
as illustrated for instance for $b=0$ in Fig.~7 of \cite{bfm1}.
When $a$ is varied, while $b$ and the horizon radius $x_h$ are kept fixed,
magnetically charged non-abelian black holes
also approach critical solutions \cite{lnw,bfm1,bfm2}.

Extending the above analysis to the case of black holes,
we observe again two regimes of $b$, 
each with its own type of critical solution corresponding to
NA-type and RN-type behaviour, respectively.
Since the transition value $a_{\rm tr} = 1.5$
is obtained by performing a Taylor expansion about a point ``far'' from the
horizon and by making use of the first few terms of this expansion,
we expect this result to hold also in the case of black holes.

Our numerical results for black holes with horizon radius $x_h=0.01$ and 0.1
are included in Figs.~2-4, along with the corresponding results
for the regular solutions.
We observe a complete analogy of the results
for regular solutions and black holes.
With increasing horizon radius $x_h$,
the transition occurs for decreasing values of $b$.
Extrapolating the curves of Fig.~2 to $a_{\rm tr}=1.5$,
we find $b_{\rm tr}\approx 24$ for $x_h=0.01$ and
$b_{\rm tr}\approx 15$ for $x_h=0.1$ \cite{foot2}.

Finally, in Fig.~6 we present $x_*/x_0$ 
as a function of the horizon radius $x_h$ for
$b = 12.5$, $b =21.125$, $b=32$ and $b = 50$.
For $b = 12.5$ and $b =21.125$ the 
monopole solutions show RN-type behaviour.
This behaviour extends to
the black hole solutions for $b=12.5$ up to $x_h \le 0.17$ and
for $b=21.125$ up to $x_h \le 0.07$.
The precise transition value must be extrapolated \cite{foot3}.
For larger values of $x_h$ NA-type behaviour arises.
For $b=32$ and $b = 50$ regular and black holes solutions show 
NA-type behaviour.
(For $b=105$ and $x_h=0.1$ only the second minimum is left,
as compared to $b \ge 400$ for the regular solutions \cite{lw}.)
For larger values of $b$ and horizon radius $x_h > 0.3$ 
the analysis of black holes solutions 
needs special consideration \cite{bfm2}.

\section{Conclusion}

For gravitating monopoles and non-abelian black holes
we have studied numerically the transition between
the low and the high Higgs mass regime.
Considering the approach of the solutions towards
a limiting critical solution,
the low Higgs mass regime shows RN-type behaviour,
i.e.~the critical solution where a horizon first appears
corresponds to an extremal RN solution
outside the horizon, while it is non-singular inside.
In contrast, the high Higgs mass regime shows NA-type behaviour,
i.e.~the critical solution is an extremal black hole with non-abelian hair.
In particular, analytical analysis suggests that
RN-type behaviour should not occur for $a < a_{\rm tr} = 1.5$
corresponding to $b > b_{\rm tr} \approx 25.6$
\cite{lw}.

Our numerical analysis indicates two main results:
\begin{itemize}
\item For asymptotically flat gravitating monopoles NA-type behaviour
is still present at $b= 30$, considerably below the numerical
value given by Lue and Weinberg \cite{lw}.
We strongly expect that the NA-type behaviour persists
until the analytical transition value is reached.
But our numerical accuracy is insufficient to show this
other than by extrapolation.

\item The analogous phenomenon occurs for black hole solutions.
Here with increasing values of the horizon radius $x_h$,
the transition occurs for decreasing values of $b$.
\end{itemize} 

\section{Appendix}

We employ a collocation method for boundary-value ordinary
differential equations developed by Ascher, Christiansen and Russell
\cite{COLSYS}.
The set of non-linear coupled differential equations eqs.~(\ref{eqmu})
-(\ref{eqH})
is solved using the damped 
Newton method of quasi-linearization.
At each iteration step a linearized problem 
is solved by using a spline collocation at Gaussian points. 
Since the Newton method works very well,
when the initial approximate solution is close to the true solution,
the gravitating monopole solutions for varying parameters
$a$ and $b$ are obtained by continuation.

The linearized problem is solved on a sequence of meshes
until the required accuracy is reached.
For a particular mesh
$x_i=x_{1} < x_{2} <...< x_{N+1}=x_o$,
where $x_i$ and $x_o$ are the boundaries of the interval,
and $h_i = x_{i+1}-x_i$, 
$h= {\rm max}_{1\le i\le N} h_i$,
a collocation solution $\vec{v}$ $(x)=(v_{1},v_{2},...v_{d})$
is determined. Each component $v_{n}(x)\in C^{m_{n}-1}[x_i,x_o]$,
is a polynomial of degree smaller than $k+m_{n}$,
where $m_{n}$ is the order of the $n$-th equation,
and $k$ is an integer bigger than the highest order
of any of the differential equations.
The collocation solution is required
to satisfy the set of differential equations
at the $k$ Gauss-Legendre points in each subinterval
as well as the set of boundary conditions.

When approximating the true solution $u_{n}(x)$
by the collocation solution $v_{n}(x)$,
an error estimate in each subinterval $x \in [x_{i},x_{i+1})$
is obtained from the expression
\begin{equation}
|| u_{n}^{(l)}(x)-v_{n}^{(l)}(x)|| _{(i)}=c_{n,l}
| u_{n}^{(k+m_{n})}(x_{i})| h_{i}^{k+m_{n}-l}+O(h^{k+m_{n}-l+1})
\nonumber \end{equation}
\begin{equation}
l=0, ...., m_{n}-1 \ , \ \ \ n=1,...,d \ , 
\label{error} \end{equation}
where $c_{n,l}$ are known constants.
Also, using eq.~(\ref{error}) a redistribution of the mesh points is performed
to roughly equidistribute the error.
With this
adaptive mesh selection procedure, the equations are solved on a sequence of
meshes until the successful stopping criterion is reached,
where the deviation of the collocation solution from the
true solution is below a prescribed error tolerance \cite{COLSYS}.

For the gravitating monopole solutions,
we typically specified 
the error tolerance in the range $10^{-4}-10^{-6}$, 
but the absolute errors of the collocation solutions
normally turned out to be far better, 
e.g.~in the range $10^{-7}-10^{-12}$. 
The number of mesh points used in these calculations
was typically about $350$,
with nearly $70\%$ of the mesh points
in the critical region of the two minima of $N(x)$, if present.

We calculated the solutions on the finite interval $[0,x_o]$
with $x_o$ ranging from 10 to $10^4$. The solutions
are highly independent of the size of the interval, as seen in
Fig.~7, where the metric function $N(x)$ is shown
for the parameters $a=1.43596$, $b=36.125$ and three value of $x_o$.
Indeed, as specified by their boundary conditions eqs.~(\ref{bc1})-(\ref{bc2}),
the gravitating monopole solutions are asymptotically flat.

In Fig.~5, where the evolution of the two minima of the metric function $N(x)$
is shown for $b=36.125$,
the corresponding global error estimates for the function $N(x)$
are also shown.

For the black hole solutions the set of boundary conditions
eqs.~(\ref{bc4})-(\ref{bc6})
is employed, and the calculations are performed on an interval
$[x_h,x_o]$, where $x_h$ represents the horizon of the black hole.
The error tolerances and the errors are comparable to those
of the gravitating monopole solutions.

\vfill
\eject

{\bf Figure captions}

\noindent Fig.~1

The metric function $N(x)$
is presented in the NA-type regime for $b=36.125$
and $a=1.43482$, 1.43550, 1.43583 and 1.43596.

\noindent Fig.~2

The critical value $a_{\rm cr}-1$ is presented as a function of $b$
in the NA-type regime
for monopole solutions and for black hole solutions
with horizon radii $x_h = 0.01$ and 0.1.

\noindent Fig.~3

The ratio $x_*/x_0$ is presented as a function of $b$
in the NA-type regime
for monopole solutions and for black hole solutions
with horizon radii $x_h = 0.01$ and 0.1.

\noindent Fig.~4

The value of the metric function $N(x)$ at the outer minimum 
at the minimal value of $a$
where the inner minimum appears
is presented as a function of $b$,
for monopole solutions and for black hole solutions
with horizon radii $x_h = 0.01$ and 0.1.

\noindent Fig.~5

The values of the metric function $N(x)$ at the outer minimum 
and at the inner minimum
are presented as a function of $a$ close to $a_{\rm cr}$
for $b=36.125$.
Also shown is the global error estimate for $N(x)$.

\noindent Fig.~6

The ratio $x_*/x_0$ is presented for black hole solutions
as a function of $x_h$
in the NA-type regime
for the values $b=12.5$, 21.125, 32, and 50,
corresponding to
$\beta = 5, 6.5, 8, 10$.

\noindent Fig.~7

The metric function $N(x)$
is presented in the NA-type regime for $b=36.125$
and $a=1.43596$,
as obtained on the interval $[0,x_o]$ with
$x_0=10$, 1000, 10000.
(The three curves cannot be distinguished graphically.)


\begin{thebibliography}{000}

\bibitem{lnw}
 K. Lee, V.P. Nair and E.J. Weinberg,
 Black holes in magnetic monopoles,
 Phys. Rev. D45 (1992) 2751.
\bibitem{bfm1}
 P. Breitenlohner, P. Forgacs and D. Maison,
 Gravitating monopole solutions,
 Nucl. Phys. B383 (1992) 357.
\bibitem{bfm2}
 P. Breitenlohner, P. Forgacs and D. Maison,
 Gravitating monopole solutions II,
 Nucl. Phys. B442 (1995) 126.
\bibitem{lw}
 A. Lue and E.J. Weinberg,
 Magnetic monopoles near the black hole threshold,
 Phys. Rev. D60 (1999) 084025.
\bibitem{bhk}
 Y. Brihaye, B. Hartmann, J. Kunz,
 Gravitating dyons and dyonic black holes,
 Phys. Lett. B441 (1998) 77.
\bibitem{bhkt}
 Y. Brihaye, B. Hartmann, J. Kunz, and N. Tell,
 Dyonic non-abelian black holes,
 Phys. Rev. D60 (1999) 104016.
\bibitem{foot}
 We obtained a decent fit for the lower $b$ points with 
     $\log N_{\rm min} = -1.5/(x-26.7)^{0.37}$.
\bibitem{lwp}
 A. Lue and E.J. Weinberg, private communication.
\bibitem{foot2}
 Extrapolation of the curves of Fig.~3 (see \cite{foot}) leads to
 $b_{\rm tr}\approx 25$ for $x_h=0.01$ and
 $b_{\rm tr}\approx 16$ for $x_h=0.1$.
\bibitem{foot3}
 Extrapolation of the upper two curves of Fig.~6 leads to the transition
 values $x_h \approx 0.14$ and $x_h \approx 0.03$ for
 $b=12.5$ and $b=21.125$, respectively.
\bibitem{COLSYS}
 U. Ascher, J. Christiansen, R.~D. Russell,
 A collocation solver for mixed order systems of boundary value problems,
 Mathematics of Computation 33 (1979) 659;\\
 U. Ascher, J. Christiansen, R.~D. Russell,
 Collocation software for boundary-value ODEs,
 ACM Transactions 7 (1981) 209.

\end{thebibliography}
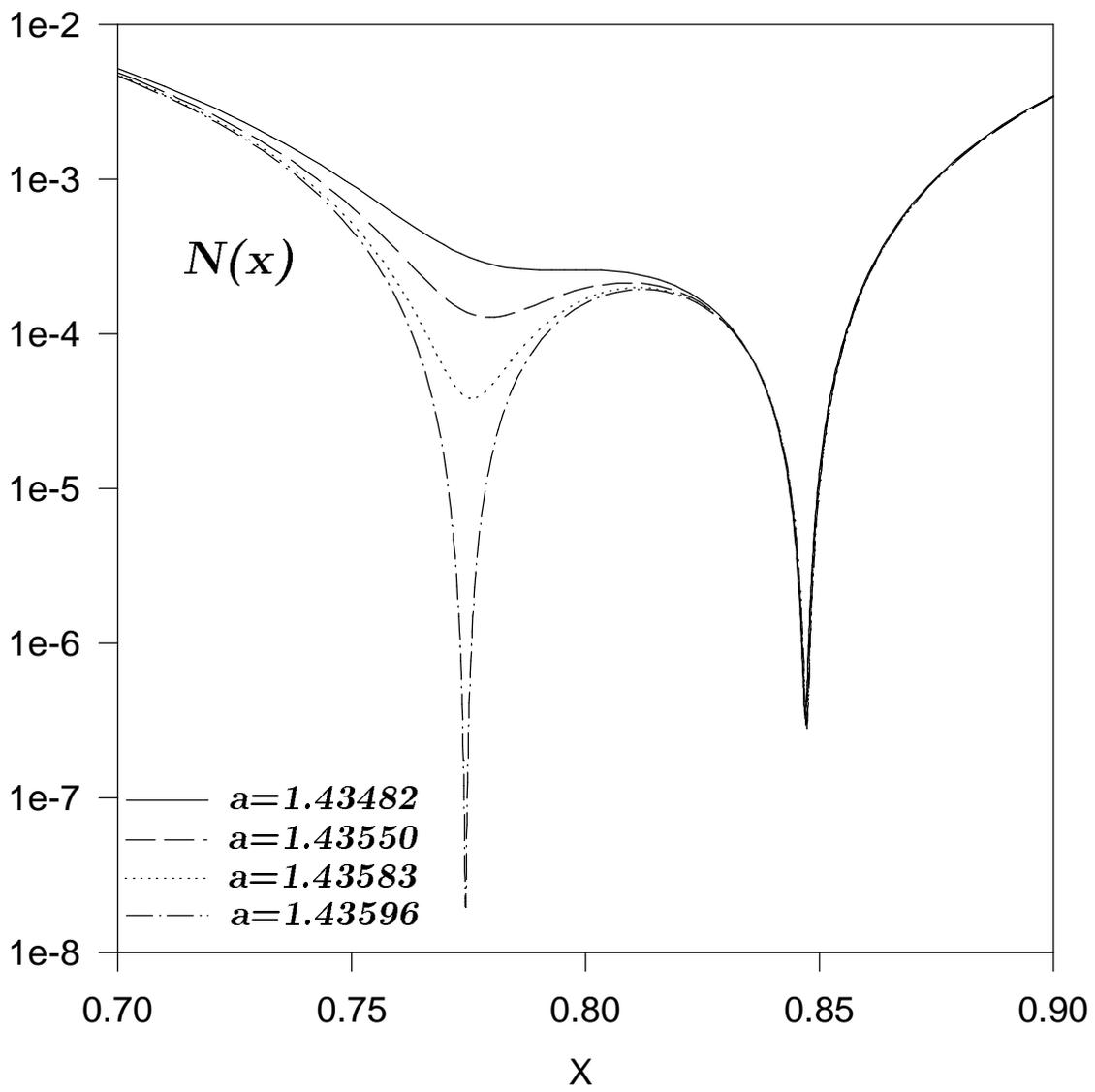
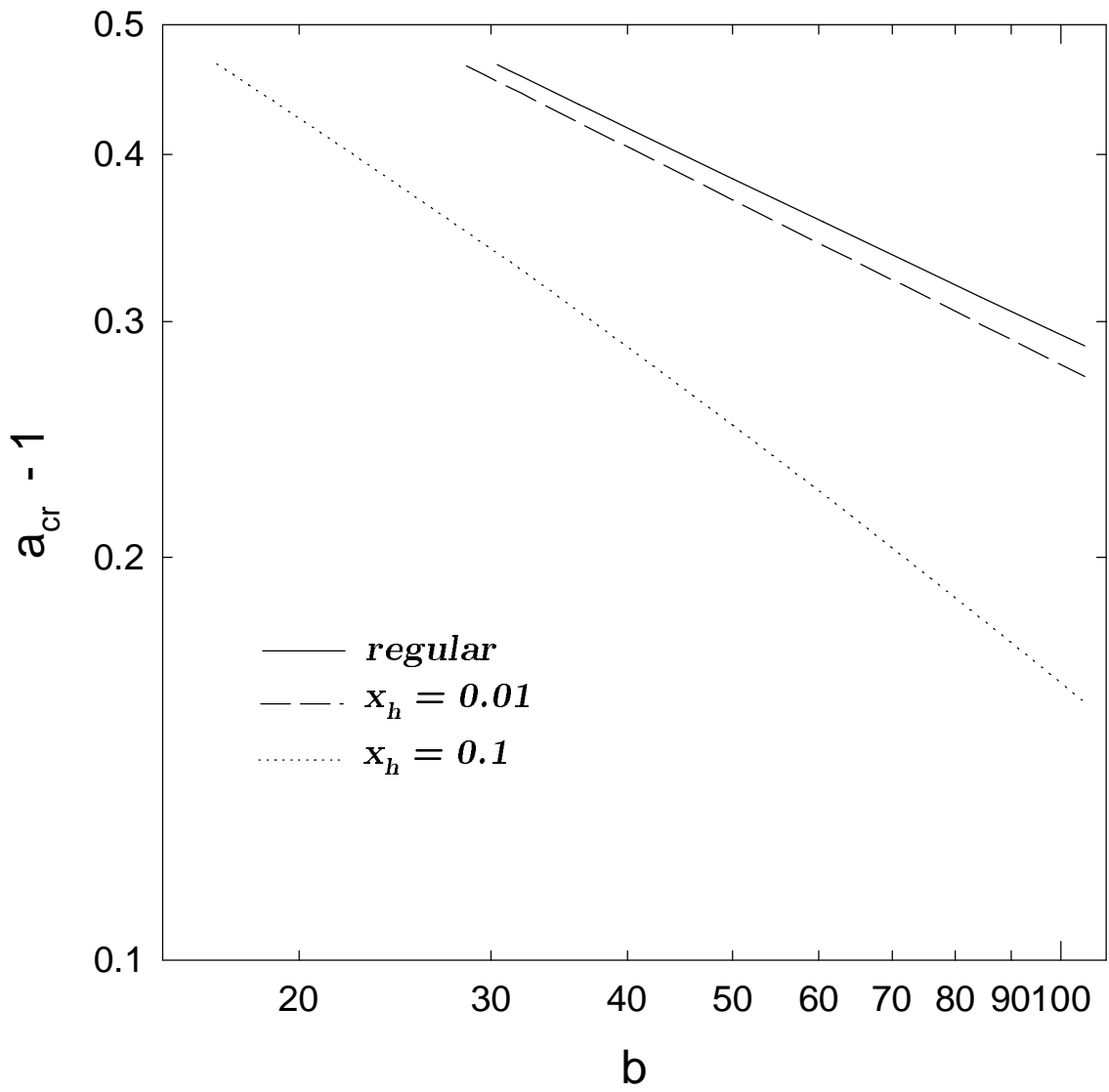
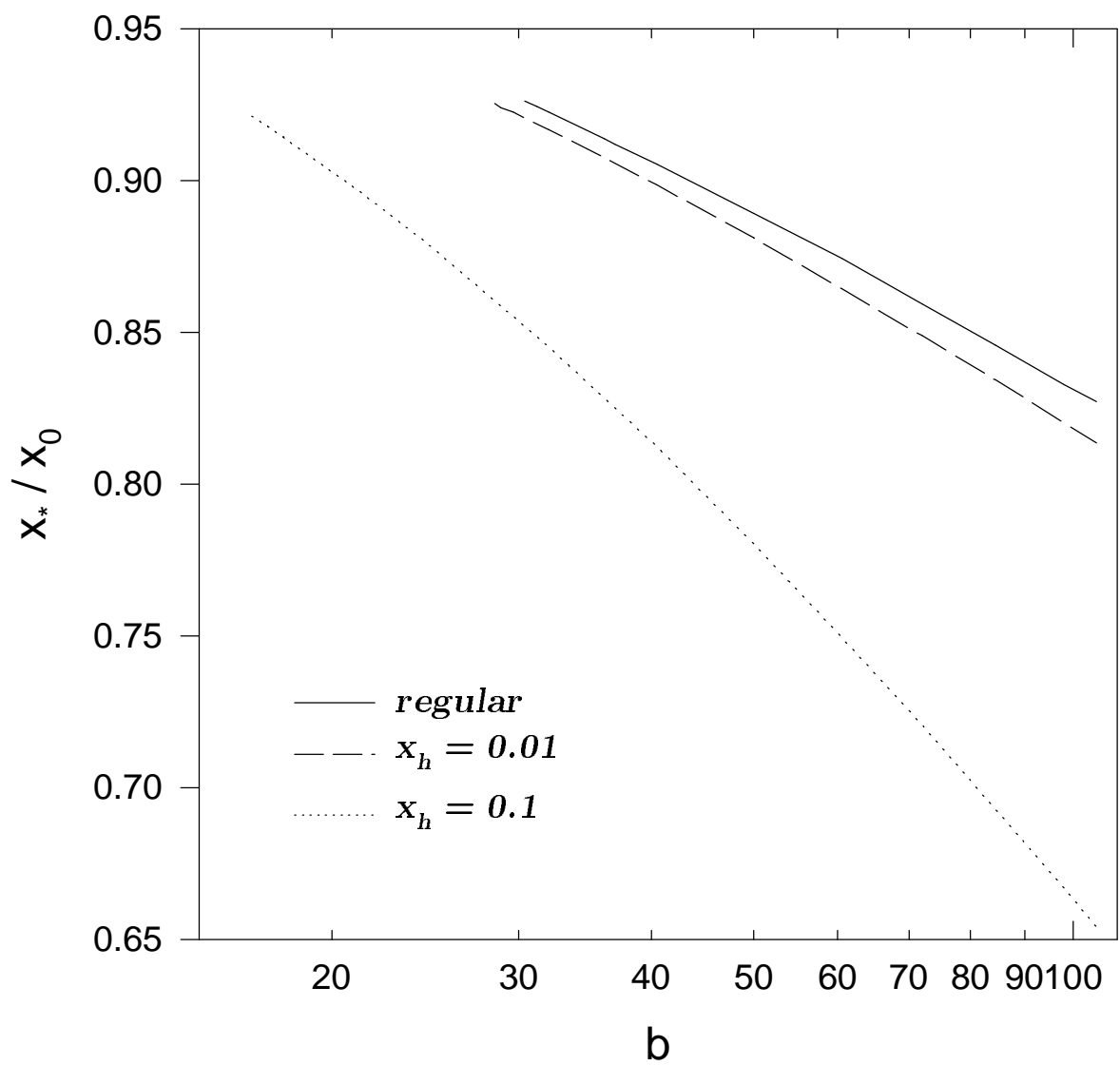
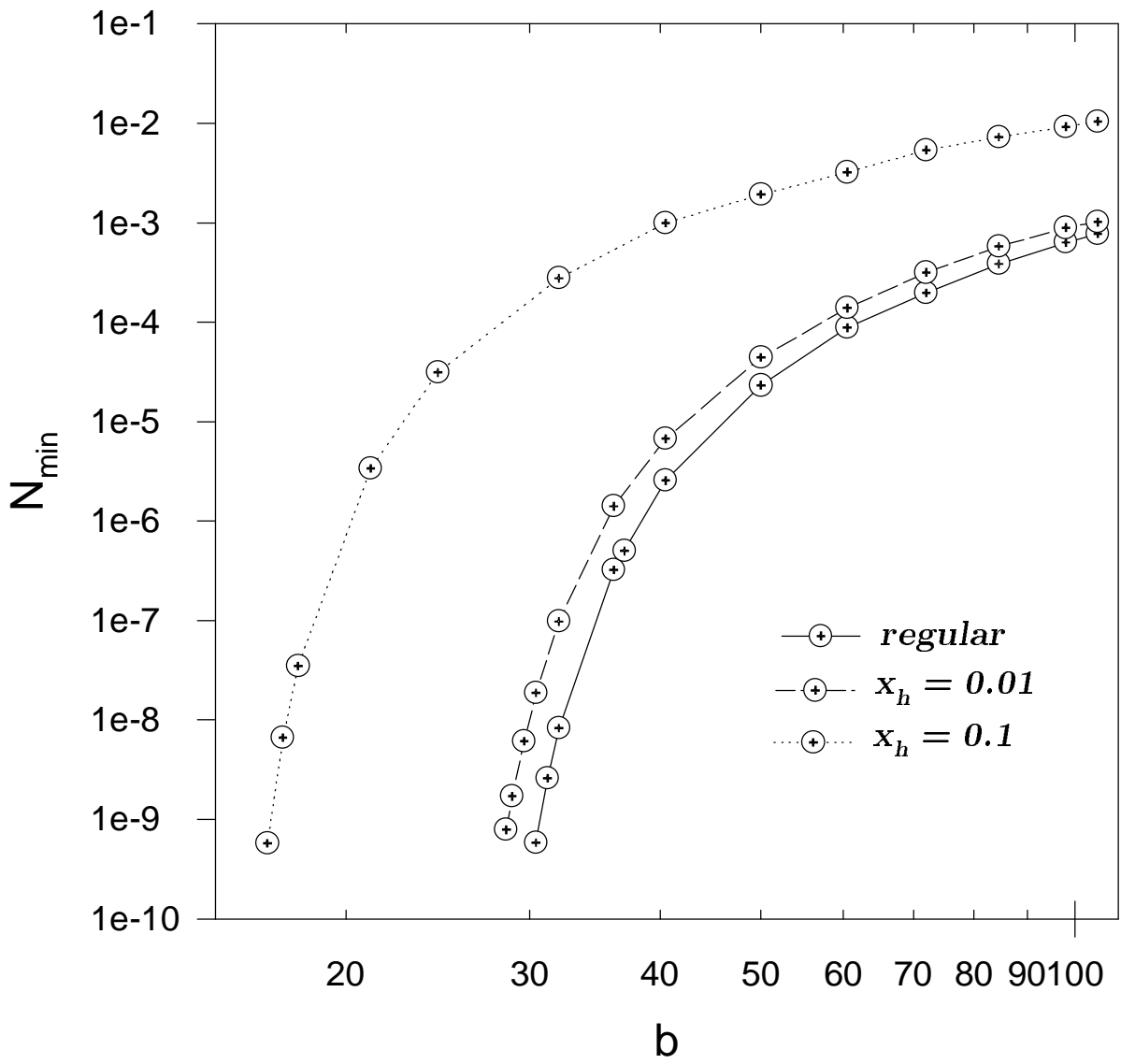
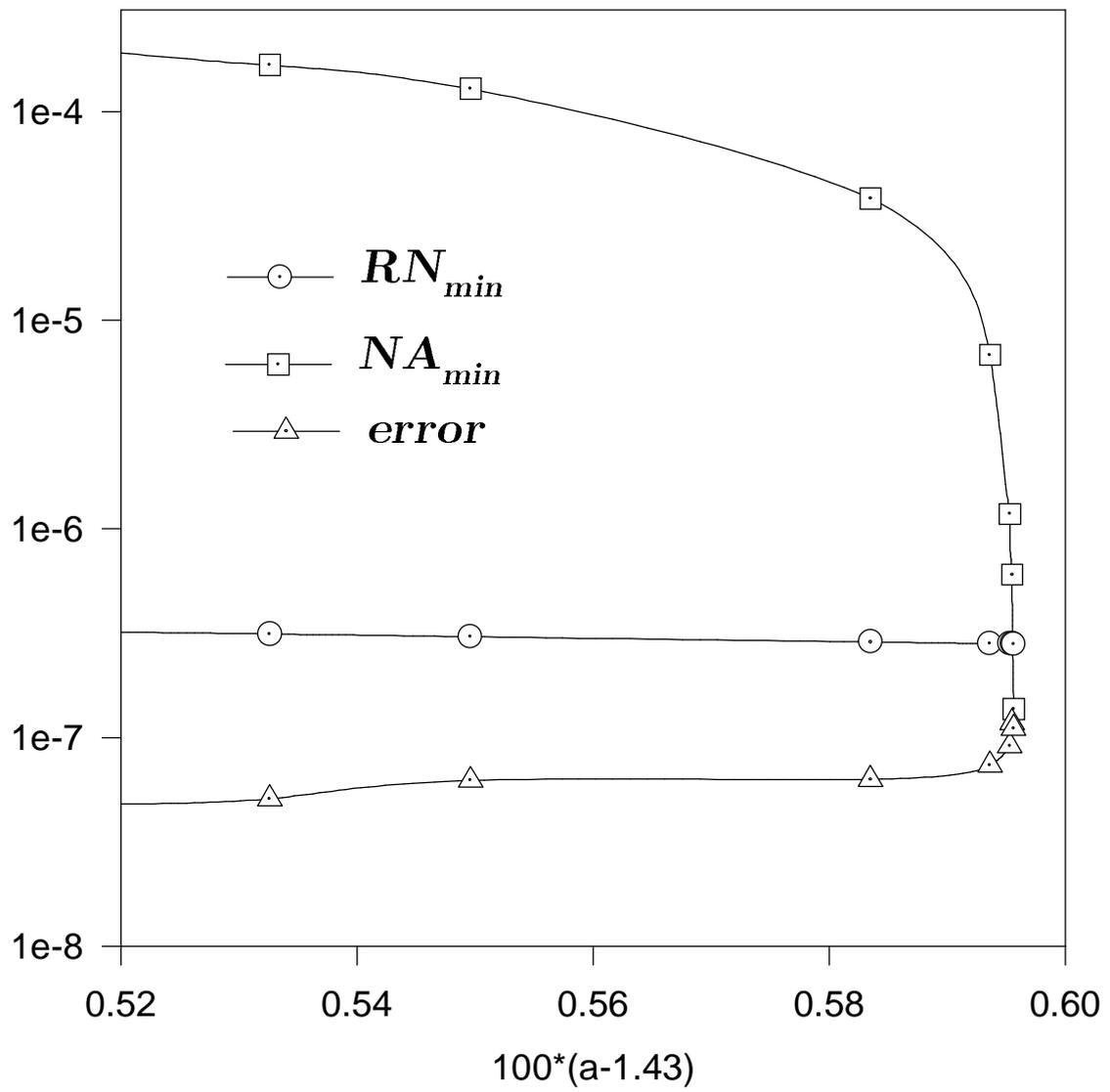
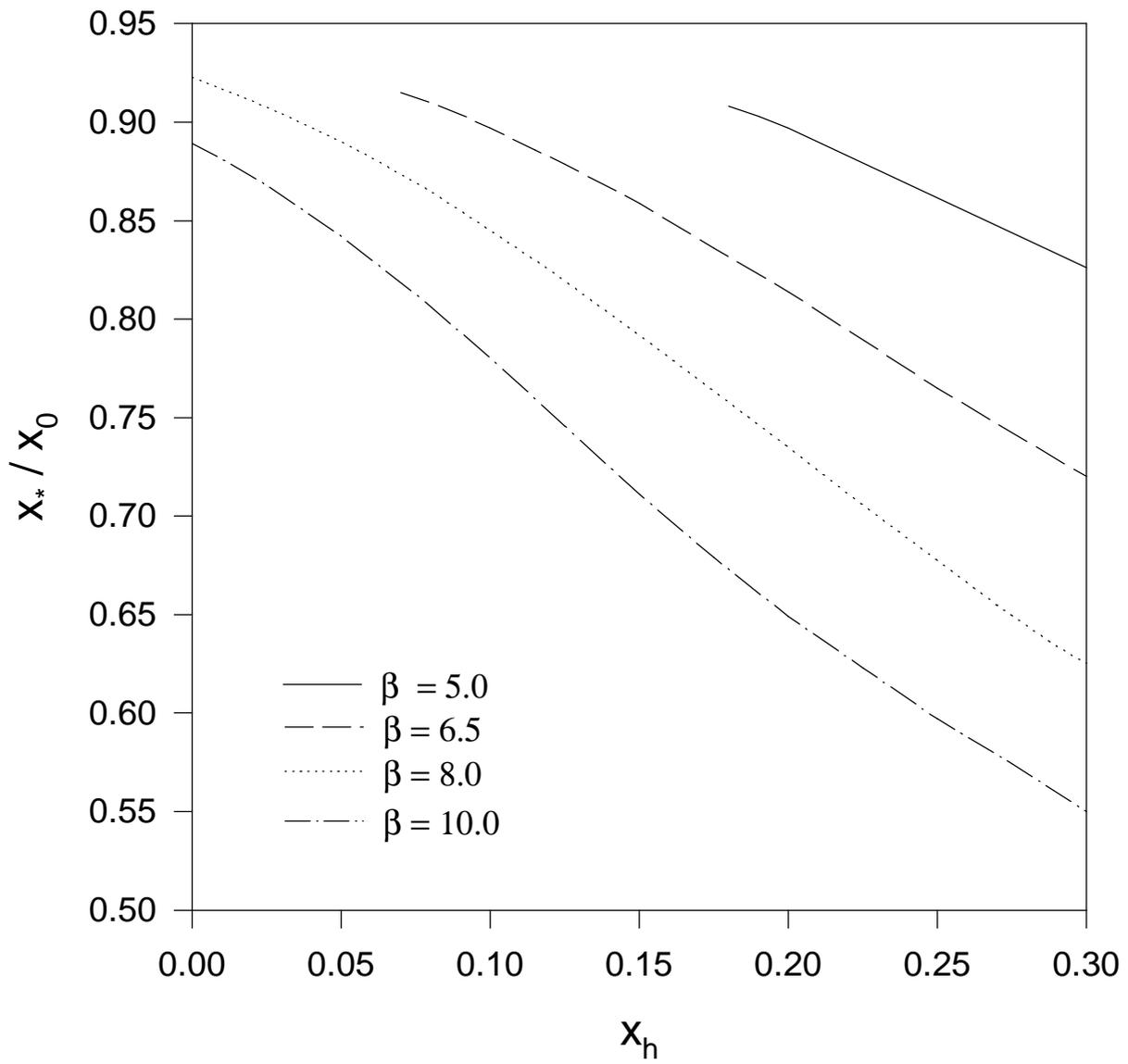
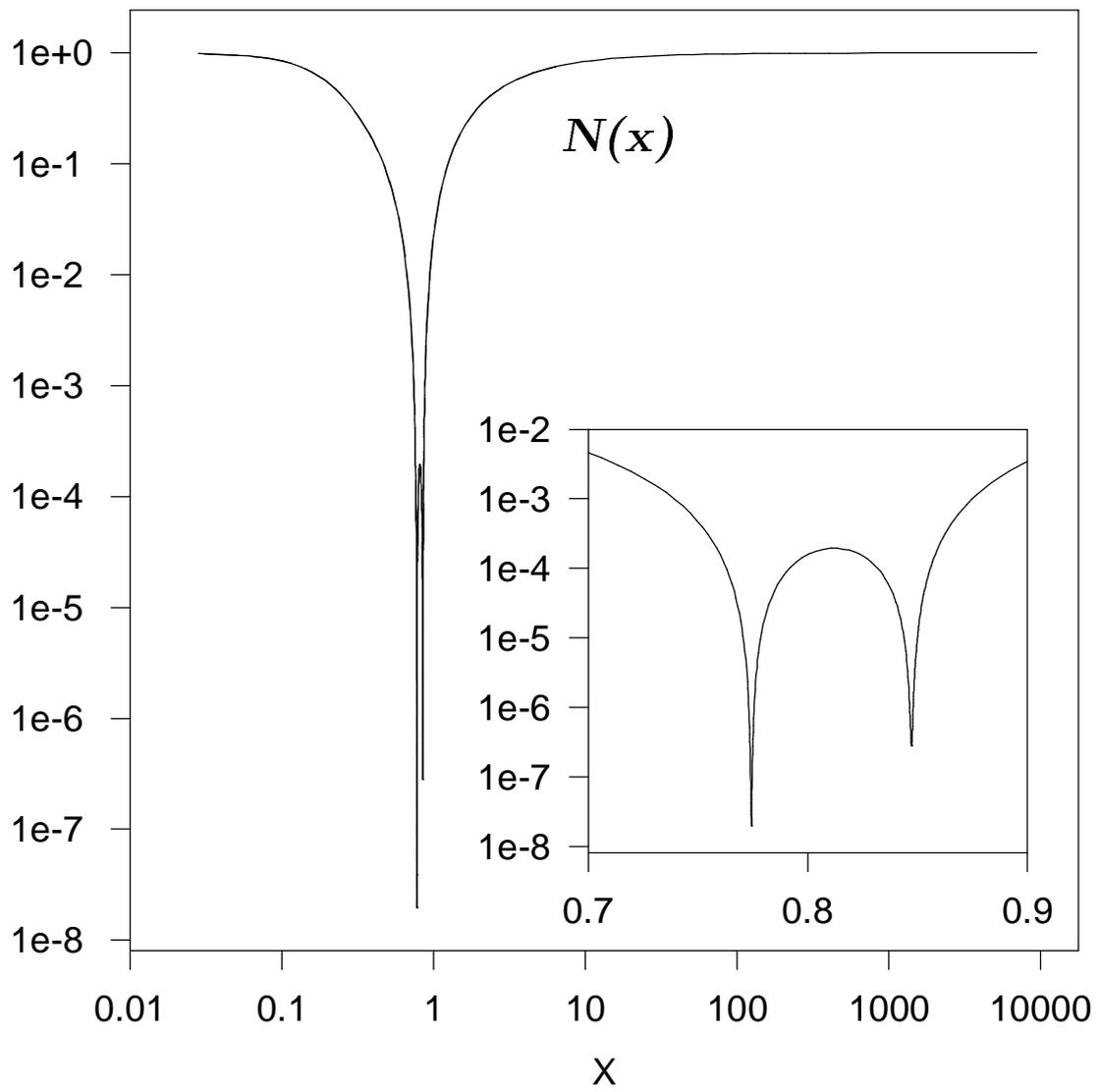
\end{document}